\documentclass[preprint,12pt]{elsarticle}

\usepackage{epsfig}
\usepackage{graphics}
\usepackage{graphicx}
\usepackage{hyperref}
\usepackage{amssymb}

\biboptions{numbers,sort&compress}

\usepackage{lineno}

\usepackage{xspace}
\usepackage{color}

\journal{Nucl. Instru. Meth. A}

\begin{document}

\def\Journal#1#2#3#4{{#1} {\bf #2}, (#3) #4}

\def\NCA{Nuovo Cimento}
\def\NIM{Nucl. Instr. Meth.}
\def\NIMA{{Nucl. Instr. Meth.} A}
\def\NPB{{Nucl. Phys.} B}
\def\NPA{{Nucl. Phys.} A}
\def\PLB{{Phys. Lett.}  B}
\def\PRL{Phys. Rev. Lett.}
\def\PRC{{Phys. Rev.} C}
\def\PRD{{Phys. Rev.} D}
\def\ZPC{{Z. Phys.} C}
\def\JPG{{J. Phys.} G}
\def\CPC{Comput. Phys. Commun.}
\def\EPJ{{Eur. Phys. J.} C}
\def\PR{Phys. Rept.}
\def\JHEP{JHEP}

\def\pt{$p_{\rm T}$\xspace}
\def\dedx{$dE/dx$\xspace}

\begin{frontmatter}



\title{Cosmic Ray Test of Mini-drift Thick Gas Electron Multiplier Chamber for Transition Radiation Detector}

\author[1,2,7]{S.~Yang\corref{cor1}}
\cortext[cor1]{Corresponding author, email address: syang@rcf.rhic.bnl.gov}
\author[3]{S.~Das}
\author[5]{B. Buck}
\author[1,7]{C.~Li}
\author[2]{T.~Ljubicic}
\author[6]{R.~Majka}
\author[1,7]{M.~Shao}
\author[6]{N.~Smirnov}
\author[4]{G.~Visser}
\author[2]{Z.~Xu}
\author[1,7]{Y.~Zhou}
\address[1]{University of Science and Technology of China, Hefei 230026, China}
\address[2]{Brookhaven National Laboratory, Upton, New York 11973, USA}
\address[3]{Institute of Physics, Bhubaneswar 751005, India}
\address[4]{Indiana University, Bloomington, Indiana 47408, USA}
\address[5]{Massachusetts Institute of Technology, Cambridge, MA 02139-4307, USA}
\address[6]{Yale University, New Haven, Connecticut 06520, USA}
\address[7]{State Key Laboratory of Particle Detection and Electronics (USTC $\&$ IHEP), USTC, Hefei 230026, China}

\begin{abstract}
A thick gas electron multiplier (THGEM) chamber with an effective readout area of 10$\times$10 cm$^{2}$  and a 11.3 mm ionization gap has been tested along with two regular gas electron multiplier (GEM) chambers in a cosmic ray test system. The thick ionization gap makes the THGEM chamber a  mini-drift chamber. This kind mini-drift THGEM chamber is proposed as part of a transition radiation detector (TRD) for identifying electrons at an Electron Ion Collider (EIC) experiment. Through this cosmic ray test, an efficiency larger than 94$\%$ and a spatial resolution $\sim$220 $\mu$m are achieved for the THGEM chamber at -3.65 kV. Thanks to its outstanding spatial resolution and thick ionization gap, the THGEM chamber shows excellent track reconstruction capability. The gain uniformity and stability of the THGEM chamber are also presented. 
\end{abstract}

\end{frontmatter}


\section{Introduction}\label{intro}
\begin{figure}[htbp]
\begin{center}
\includegraphics[keepaspectratio,width=0.9\textwidth]{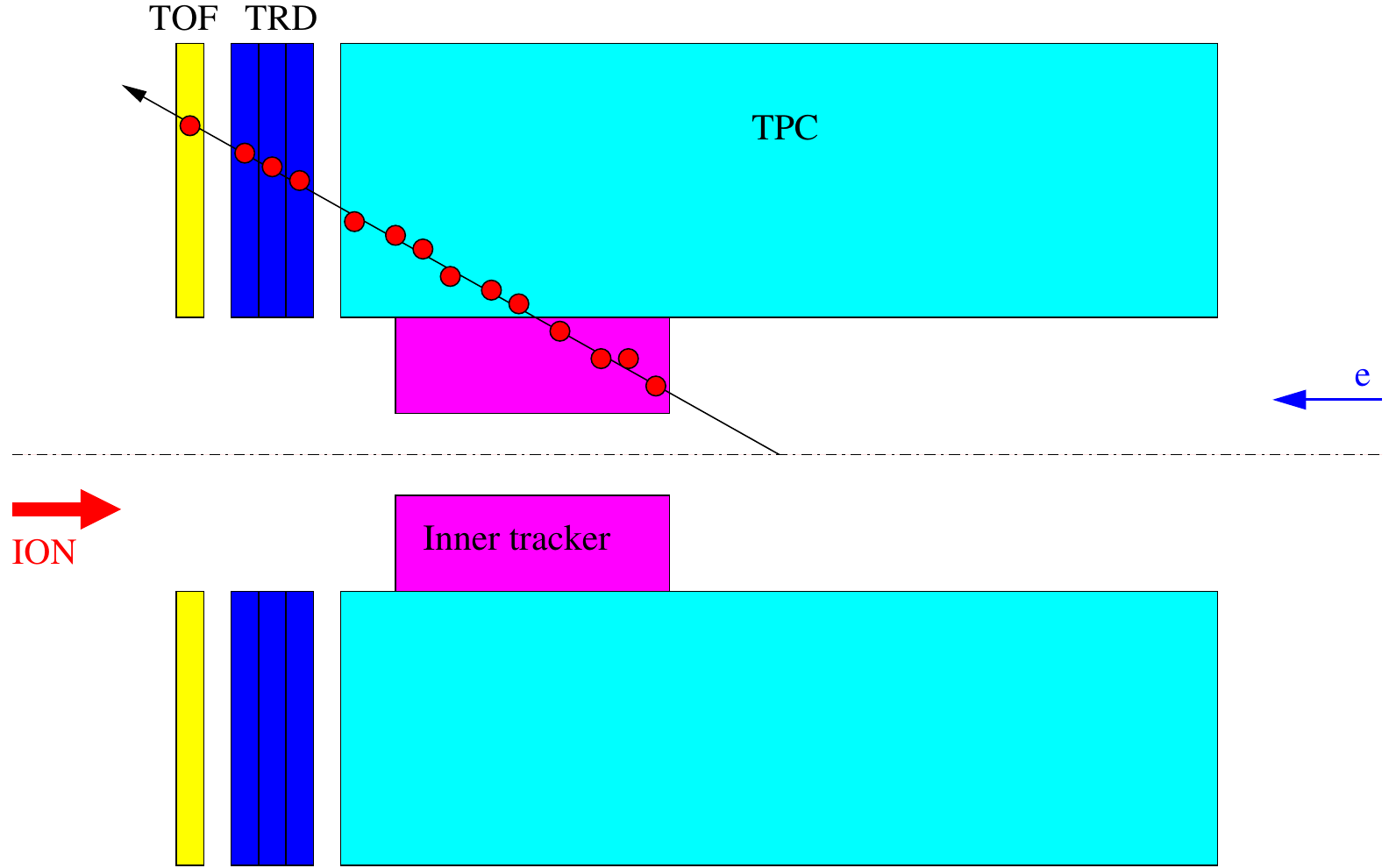}
\vspace*{0mm}
\caption{A schematic view of the detector configuration at eSTAR. The proposed TRD+eTOF is placed between the pole-tip and the low-material tracking detectors.} \label{trd_setup}
\end{center}
\end{figure}

\begin{figure}[htbp]
\begin{center}
\includegraphics[keepaspectratio,width=0.7\textwidth]{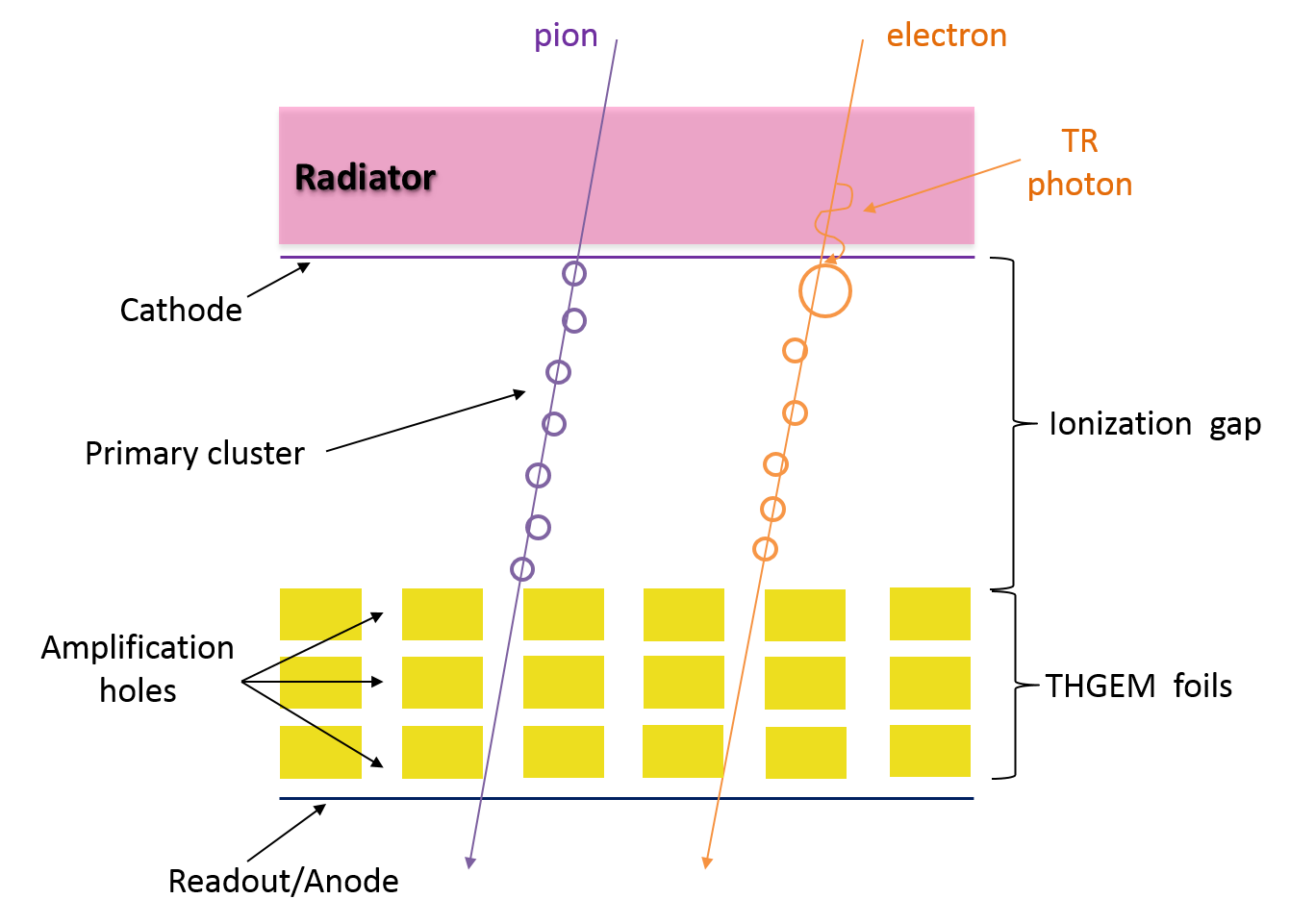}
\vspace*{-4mm}
\caption{Schematic of the THGEM based TRD.} \label{trd_working_principle}
\end{center}
\end{figure}

An Electron Ion Collider (EIC) is being considered as the next generation QCD facility to understand how the visible universe is built up~\cite{eic}. More specifically, the EIC will probe with unprecedented precision the low Bjorken-x domain where gluons and sea quarks dominate for both nucleons and nuclei. A possible realization of the  accelerator facility based on the (currently operating) Relativistic Heavy Ion Collider (RHIC), called eRHIC, has been proposed~\cite{erhic}. The Solenoidal Tracker at RHIC (STAR) detector, one of the two major experiments at RHIC, has been planned to evolve into eSTAR with a suite of upgrades optimized for the EIC physics program. The eSTAR detector performance and a broad range of flagship measurements, which have been identified as part of the EIC science case, have been studied through simulation. eSTAR has been found to be well suitable for an initial stage of eRHIC~\cite{estarloi}. However, one of the major experimental challenges is to cleanly identify the scattered electron and to provide precise kinematics of the interaction. We have proposed a compact transition radiation detector (TRD) followed by an endcap time-of-flight (eTOF) detector with an additional converter~\cite{startrdproposal}. Figure~\ref{trd_setup} illustrates a schematic view of the detector configuration relevant to this proposal. The low-material time projection chamber (TPC) (and a possible inner tracker) in a solenoidal field in front of TRD, provides tracking, momentum, \dedx and photon rejection. The combination of eTOF and TPC provides electron identification at electron momentum less than a few GeV/$c$~\cite{tpctofpid}. Furthermore, eTOF provides hadron identification in the case that both hadrons and the scattered electrons strike the eTOF, as well as the collision time reference. The proposed TRD is similar to the ALICE TRD~\cite{alicetrdproposal}, providing \dedx measurement in addition to the transition radiation (TR) signal and tracklet reconstruction capability, but with a readout stage based on thick gas electron multiplier (THGEM) chamber rather than the multiple wire proportional chamber (MWPC). The operation principle and design of the THGEM based TRD chamber is depicted by Fig.~\ref{trd_working_principle}. The TRD serves two functions. Firstly, it provides additional \dedx measurement with a Xe rich gas mixture at the entire momentum range. This is essential for small angle scattering, where only a small section of the particle trajectory falls within the tracking detector acceptance, resulting in few hits in the tracking detector and worse \dedx resolution (1/$\sqrt{N}$ rule). Through the simulation~\cite{startrdproposal}, TRD with 300 $\mu$m spatial resolution is enough for tracking due to the thick TRD radiator and material in the TPC endcap. Secondly, it adds necessary TR signal to particles at high momentum. The TR threshold is around $\gamma$~$>$~1000$\sim$2000. From a practical standpoint, only electrons provide such radiation into the ionization chamber in the TRD, boosting the effective electron \dedx to even higher values from the existing relativistic rise.

Comparing with the gas electron multiplier (GEM) chamber first introduced in 1996 at CERN~\cite{gem}, the THGEM chamber is one of the most recently developed micropattern gas detectors~\cite{thgem}. The THGEM is a robust, high-gain gaseous electron multiplier, and has a hole-structure similar to the GEM~\cite{thgem1}. It is manufactured economically by mechanically drilling sub-millimeter diameter holes in a thin printed-circuit board (PCB), followed by Cu-etching of the hole’s rim.

Due to lack of Xe in laboratory, we used the Ar mixture instead of Xe mixture as the working gas of the THGEM chamber. Some experiments were carried out for comparing the THGEM or GEM chamber performance in large variety of gases~\cite{arxethgem,arxethgem1,arxegem,arxegem1}. The maximum gain measured with THGEM chamber operating in Xe is similar to that measured with THGEM chamber operating in Ar at atmospheric pressure~\cite{arxethgem}. Meanwhile, the maximum GEM chamber gains measured in Ar/CO$_{2}$ and Xe/CO$_2$ are comparable at atmospheric pressure~\cite{arxegem}. Moreover, the results in~\cite{arxethgem} show the best energy resolution of the specified THGEM chamber reached in Ar with 5.9 keV x-rays is 30$\%$ FWHM while that reached in Xe is 27$\%$ FWHM. These previous experiment results illustrate the results measured with the THGEM chamber operating in Ar mixture are able to provide a valid reference for that operating in Xe mixture.

In this paper, we focus on THGEM chamber's performance in various aspects such as detection efficiency, spatial resolution, gain uniformity and stability, especially the track reconstruction capability. The paper is organized as follows. Section 2 describes the THGEM chamber. The cosmic ray test system setup is presented in Section 3. Section 4 describes the performance of the THGEM chamber. Section 5 provides a concluding summary.

\section{The THGEM chamber}\label{thgem}
\begin{figure}[htbp]
\begin{center}
\includegraphics[keepaspectratio,width=1.0\textwidth]{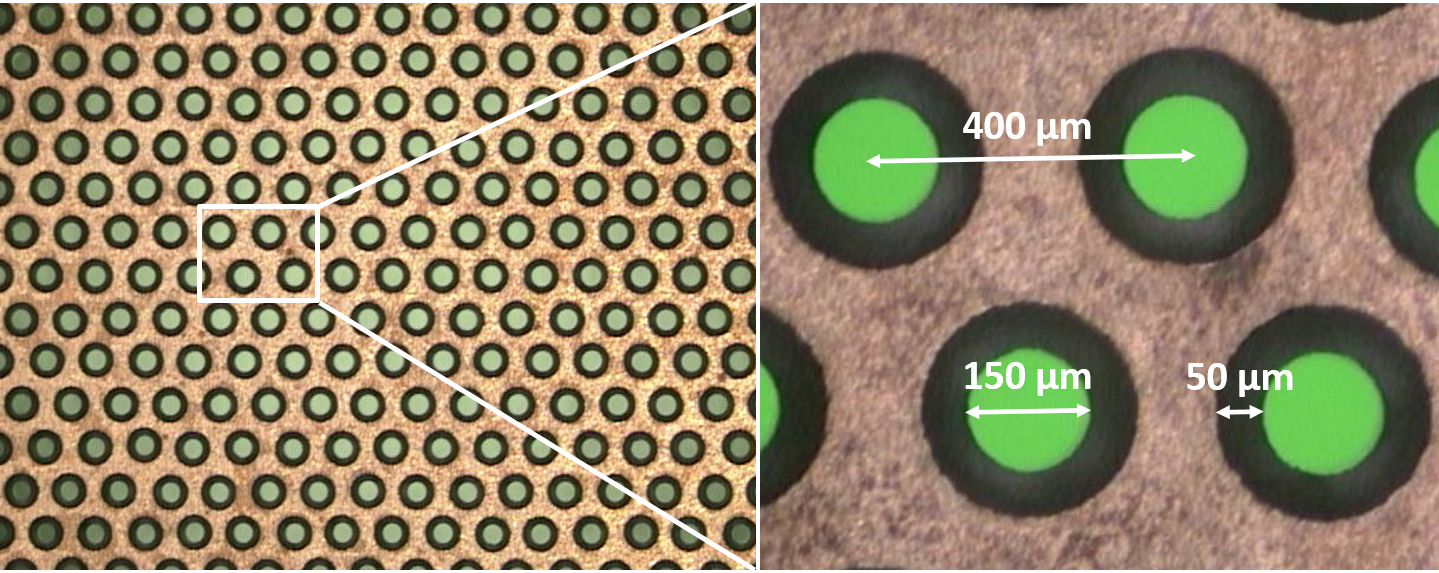}
\vspace*{-3mm}
\caption{The structure of the THGEM foil.} \label{foil_structure}
\end{center}
\end{figure}

\begin{figure}[htbp]
\begin{center}
\includegraphics[keepaspectratio,width=0.6\textwidth]{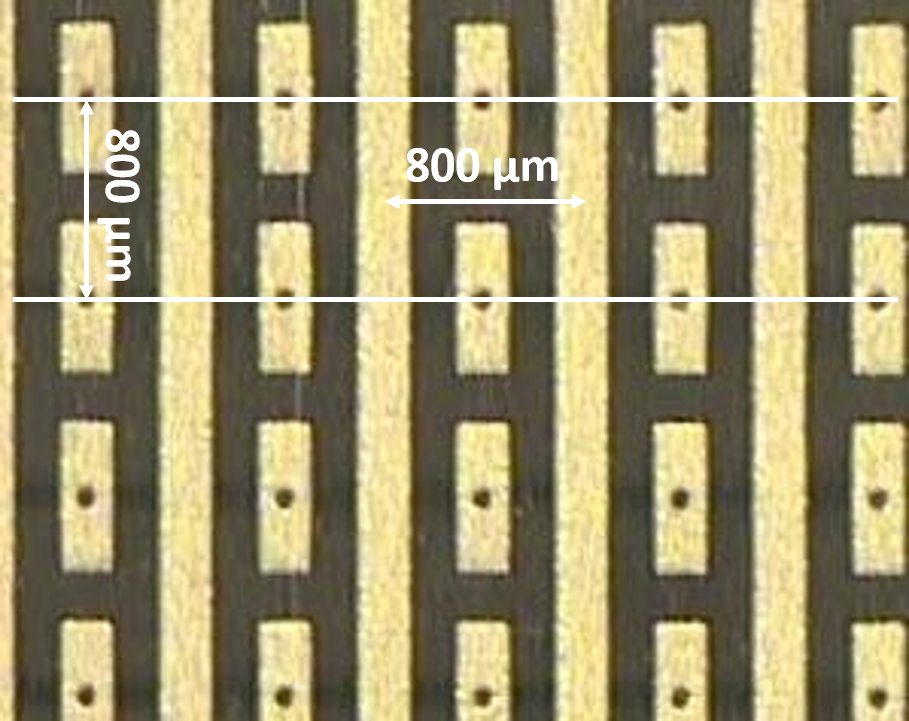}
\vspace*{+0mm}
\caption{The readout board structure of the THGEM chamber.} \label{readout_structure}
\end{center}
\end{figure}

The THGEM chamber in this study uses three THGEM foils in cascade. These foils are provided by the Institute of High Energy Physics (IHEP), China. The foil structure is shown in Fig.~\ref{foil_structure}. The major parameters include foil thickness 300 $\mu$m, hole diameter 150 $\mu$m, and hole pitch 400 $\mu$m. The rim clearance region around the hole is $\sim$50 $\mu$m. The effective readout area given by the foil  is 10$\times$10 cm$^{2}$. Figure~\ref{readout_structure} depicts the readout board structure of the THGEM chamber. The readout unit, with a pitch of 800 $\mu$m, is strip-style in $x$ direction while pad-style (inter-connected beneath) in $y$ direction. The thickness of the THGEM chamber ionization gap is 11.3 mm. The THGEM chamber is placed in a gas-tight aluminum box with a gas mixture (90$\%$ Ar + 10$\%$ CO$_{2}$) at atmospheric pressure.

\begin{figure}[htbp]
\begin{center}
\includegraphics[keepaspectratio,width=1.0\textwidth]{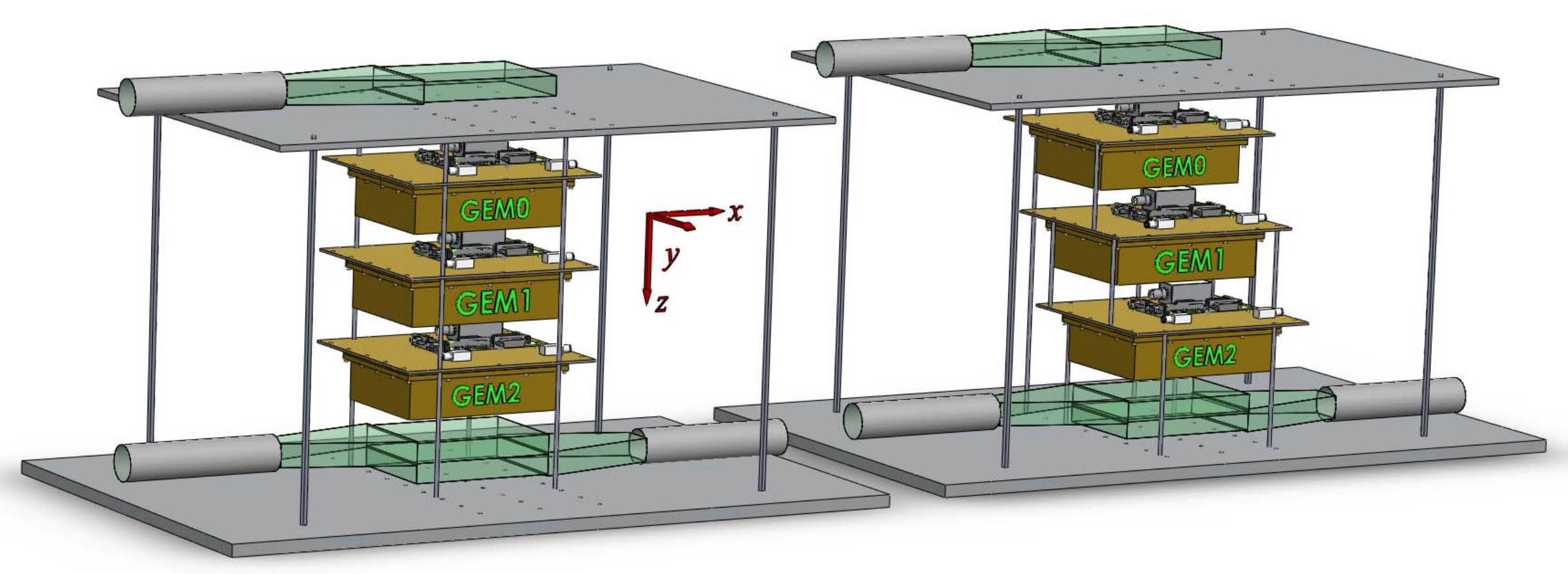}
\vspace*{-3mm}
\caption{Schematic of the two different setups of the cosmic ray test system. Left Panel: The setup with the THGEM chamber (GEM1) and two regular GEM chambers (GEM0, GEM2) aligned vertically. Right Panel: The setup with 4.1 cm horizontal offset between the THGEM chamber and the regular GEM chamber along $y$ direction. The distance between the THGEM chamber and the regular GEM chamber along $z$ direction is 10.5 cm.} \label{cosmic_system_setup}
\end{center}
\end{figure}

\begin{figure}[htbp]
\begin{center}
\includegraphics[keepaspectratio,width=1.0\textwidth]{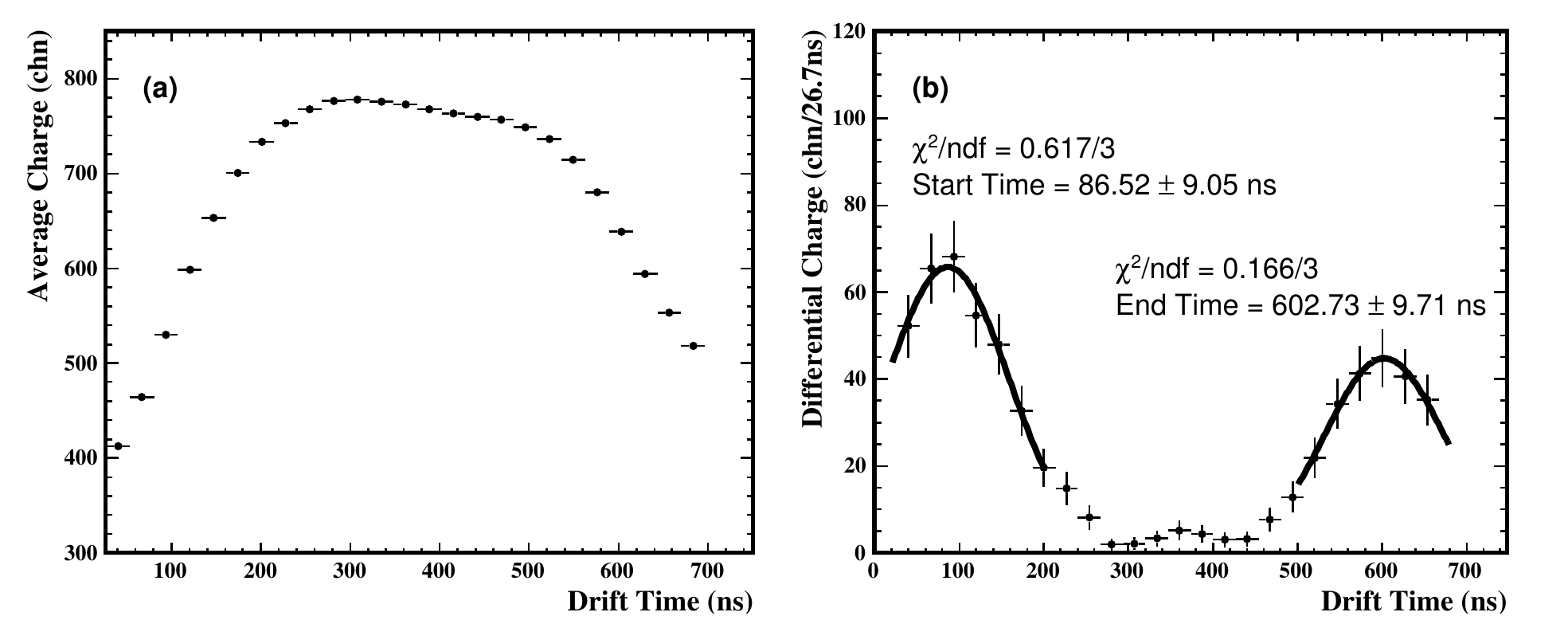}

\vspace*{-3mm}
\caption{(a) The typical average charge vs. drift time of the THGEM chamber at operating voltage. (b) The differential spectrum of the average charge vs. drift time.} \label{adcvstb}
\end{center}
\end{figure}

\section{Cosmic ray test system setup}\label{systemsetup}
Two different setups of the cosmic ray test system are shown in Fig.~\ref{cosmic_system_setup}. Three scintillators read out by photomultipliers, providing trigger for this system, are placed upstream and downstream of the three GEM chambers. The effective trigger area is around 12$\times$12 cm$^{2}$. The system consists of two regular GEM chambers (GEM0, GEM2) and one THGEM chamber (GEM1) in between. The vertical ($z$ direction) distance between the THGEM chamber and the regular GEM chamber is 10.5 cm. The readout boards of the two regular GEM chambers, with 3.8 mm ionization gap width, are identical to that of  the THGEM chamber. These two GEM chambers are used to calibrate the THGEM chamber. They can also be used to measure the cosmic ray tracklet slope to study the track reconstruction capability of the THGEM chamber. The left setup in Fig.~\ref{cosmic_system_setup} with all the detectors aligned vertically is used to measure the detection efficiency of the THGEM chamber while the right setup with 4.1 cm horizontal offset between the THGEM chamber and the regular GEM chamber along $y$ direction is used to study other performance of the THGEM chamber, such as spatial resolution, track reconstruction capability,  gain uniformity and stability. The front end electronics (FEE) of these three GEM chambers are all based on the APV25-S1 chip~\cite{apv}, and the readout system is almost the same as that of the forward GEM tracker (FGT) at the STAR experiment~\cite{fgtreadout}. The only difference is that the front end card has only 2 APV chips rather than 5 used for the FGT. Each readout unit is sampled in 27 time bins (26.7 ns bin width) along electron drift direction ($z$ direction). The typical  average charge as a function of drift time for the mini-drift THGEM chamber at operating voltage measured by this readout system is shown in Fig.~\ref{adcvstb}(a), and its width is the drift time in the ionization gap. The drift time in the ionization gap can be calculated through the differential spectrum of the average charge vs. drift time, shown in Fig.~\ref{adcvstb}(b). Thus, the electron drift velocity can be derived and the value is 2.19 cm/$\mu$s (1.13/(0.603-0.087) cm/$\mu$s). Because we used 3 THGEM foils in cascade, the effective drift length should be a little larger than the ionization gap width. So the drift velocity measured by this method is a little smaller than the real velocity value.  

\section{The performance of the THGEM chamber}\label{results}
\subsection{The efficiency plateau of the THGEM chamber}
\begin{figure}[htbp]
\begin{center}
\includegraphics[keepaspectratio,width=0.6\textwidth]{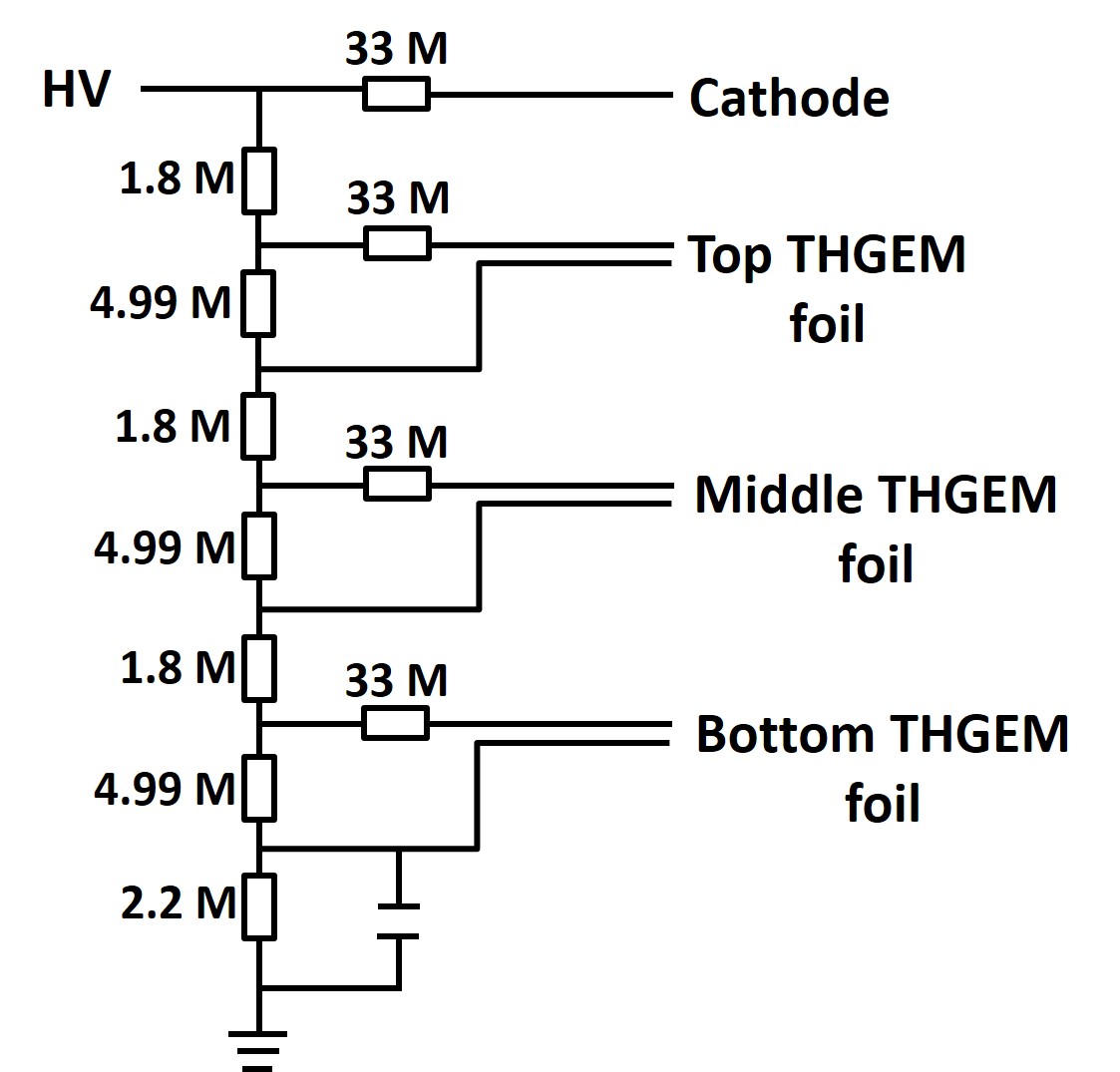}
\vspace*{-3mm}
\caption{Schematic of the THGEM chamber HV divider.} \label{hv_divider}
\end{center}
\end{figure}

\begin{figure}[htbp]
\begin{center}
\includegraphics[keepaspectratio,width=0.6\textwidth]{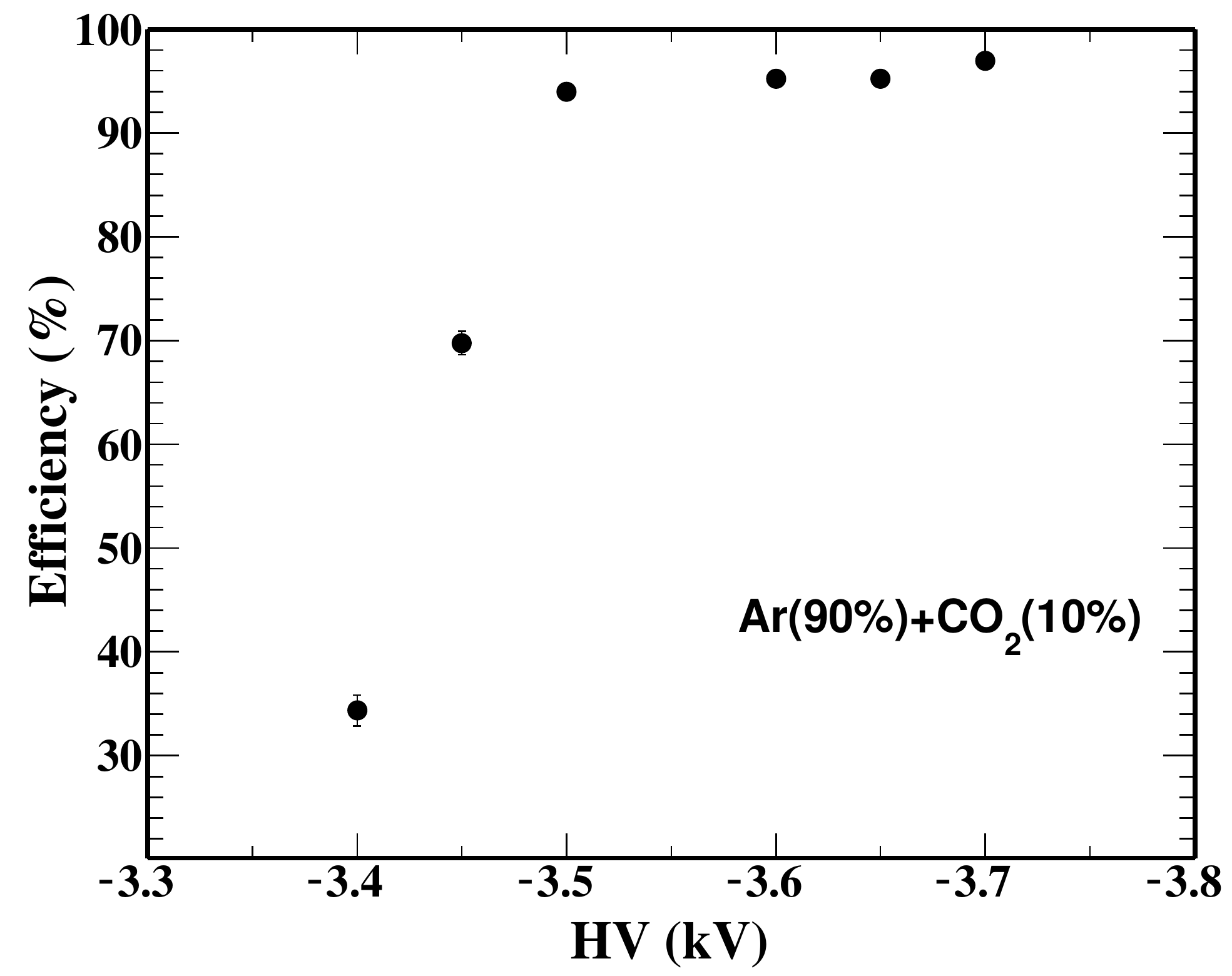}
\vspace*{-3mm}
\caption{The efficiency plateau of the THGEM chamber.} \label{eff_plateau}
\end{center}
\end{figure}

In order to find the optimum operating voltage of the THGEM chamber, the detection efficiency is scanned as a function of HV with the cosmic ray test system. A schematic view of the THGEM chamber HV divider is shown in Fig.~\ref{hv_divider}. The efficiency plateau of the THGEM chamber is shown in Fig.~\ref{eff_plateau}. The detection efficiency goes above 90$\%$ when the applied HV is higher than -3.5 kV. -3.65 kV is selected as the operating HV for the THGEM chamber at which the efficiency is greater than 94$\%$.

\subsection{The spatial resolution of  the THGEM chamber}
\begin{figure}[htbp]
\begin{center}
\includegraphics[keepaspectratio,width=1.0\textwidth]{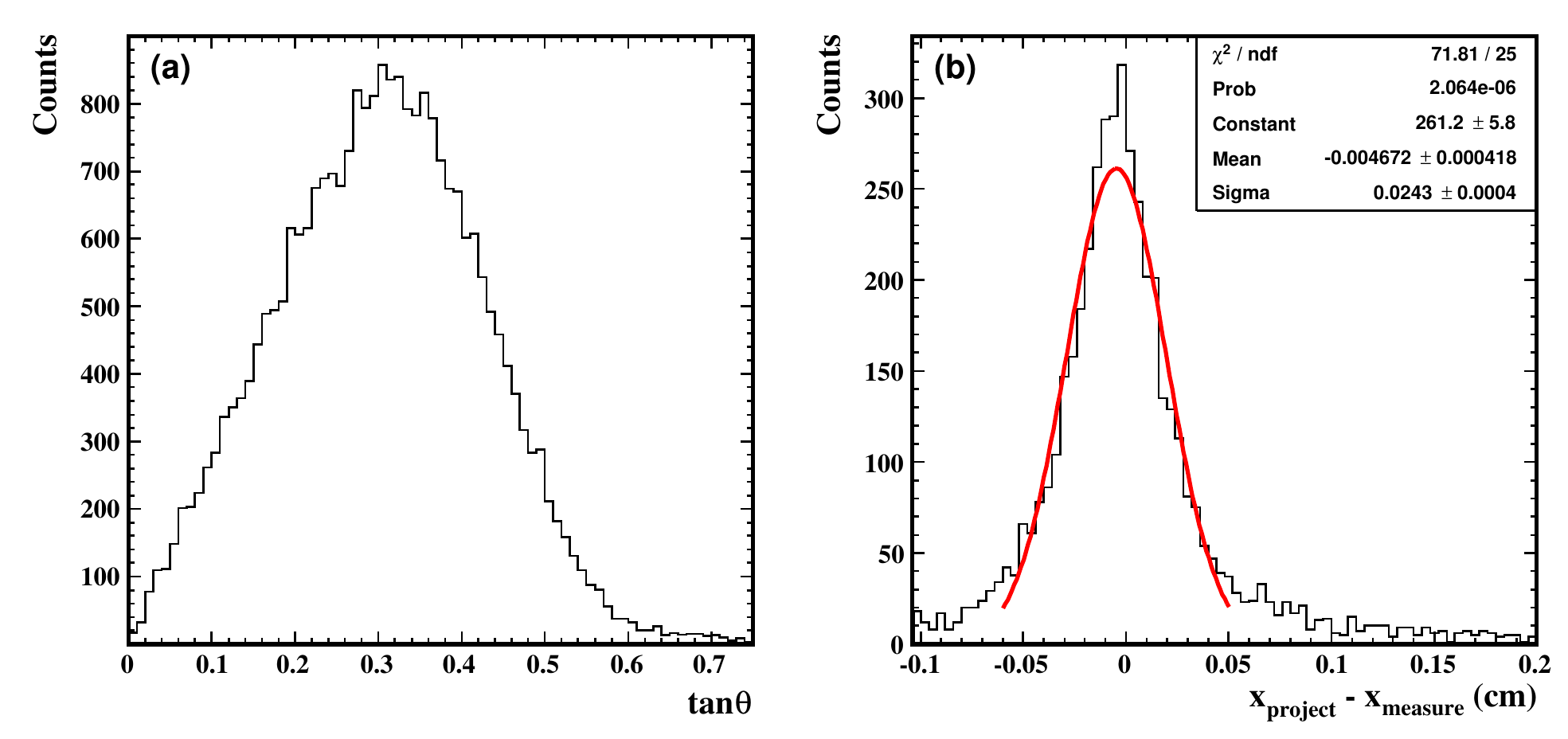}
\vspace*{-3mm}
\caption{(a) The cosmic ray zenith (incident) angle distribution. (b)  The residual distribution in $x$ direction, $x_{project}$ is the $x$ position at the THGEM chamber projected from the two regular GEM chambers, $x_{measure}$ is the $x$ position measured by the THGEM chamber using the center of gravity method.} \label{thgem_spatial}
\end{center}
\end{figure}

The two regular GEM chambers' performance was tested elsewhere before this cosmic ray test. Their spatial resolutions are found to be better than 150 $\mu$m~\cite{gemres}. The zenith angle ($\theta$) distribution of cosmic ray recorded by this system is shown in Fig.~\ref{thgem_spatial}(a). The zenith angle is calculated by the cosmic ray hit positions on the two regular GEM chambers using the following expression tan$\theta = \frac{\sqrt{(x_{0}-x_{2})^{2}+(y_{0}-y_{2})^{2}}}{|z_{0}-z_{2}|}$, where $x_{0}$, $y_{0}$ are the $x$, $y$ positions measured by GEM0 using the center of gravity method while $x_{2}$, $y_{2}$ are provided by GEM2. $z_{0}$, $z_{2}$ are the positions of GEM0 and GEM2 along $z$ direction. The vertically incident cosmic ray tracks are selected (tan$\theta$ $<$ 0.1) to measure the spatial resolution of the THGEM chamber. Figure~\ref{thgem_spatial}(b) shows the residual distribution in the $x$ direction, $x_{project}-x_{measure}$, where $x_{project}$ is the cosmic ray trajectory position at the THGEM chamber projected from the two regular GEM chambers and $x_{measure}$ is the hit position measured by the THGEM chamber. The residual distribution in $y$ direction is very similar. The 1-D spatial resolution of the THGEM chamber is $\sim$220 $\mu$m after subtracting the contribution from the two regular GEM chambers. 

\subsection{Electron drift velocity measured by the THGEM chamber}
\begin{figure}[htbp]
\begin{center}
\includegraphics[keepaspectratio,width=0.8\textwidth]{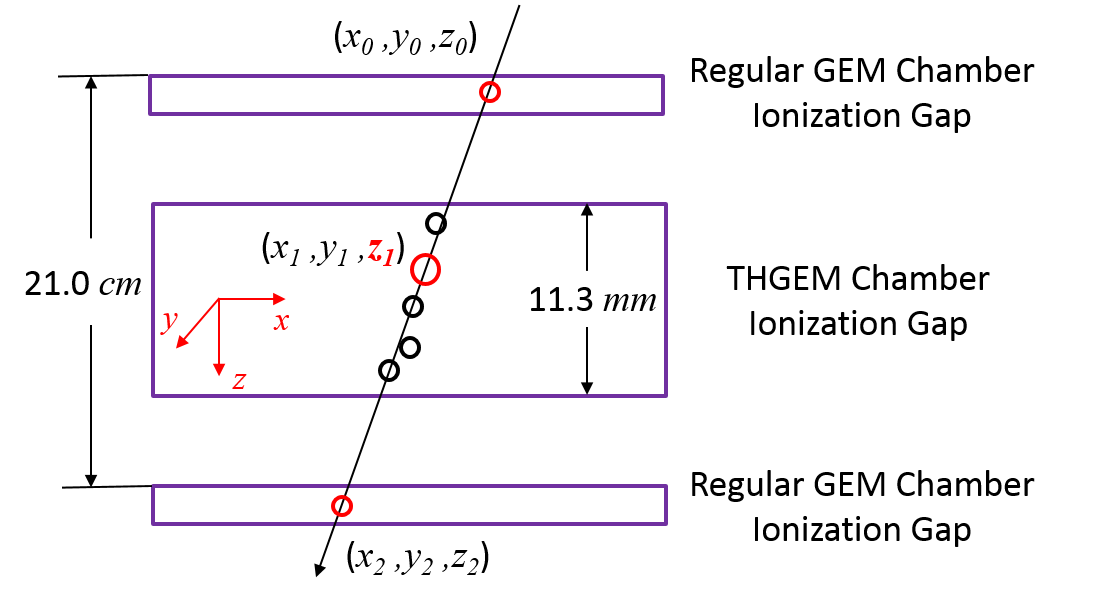}
\vspace*{-3mm}
\caption{The principle of deriving maximum energy deposit point's z position.} \label{driftv_principle}
\end{center}
\end{figure}

\begin{figure}[htbp]
\begin{center}
\includegraphics[keepaspectratio,width=0.98\textwidth]{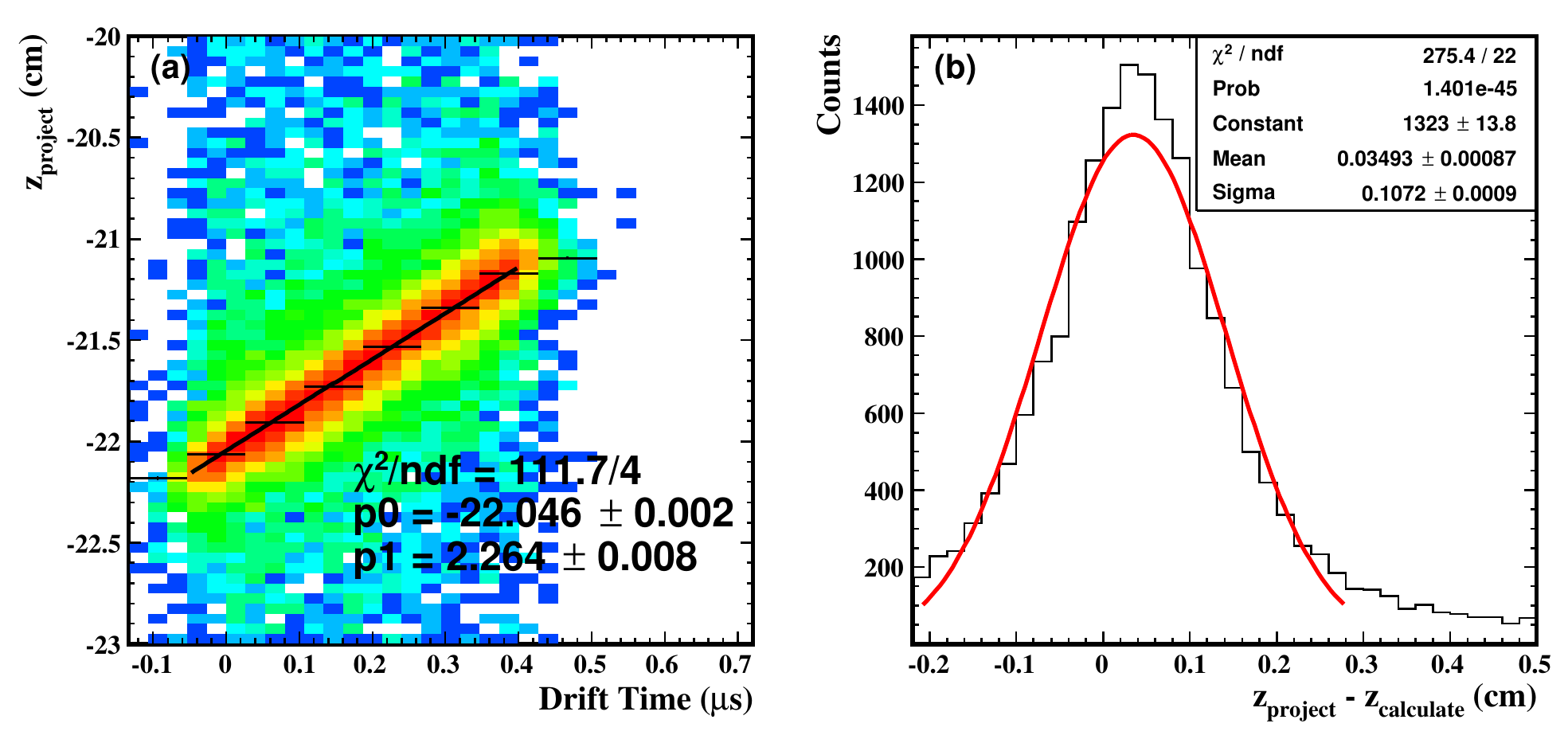}
\vspace*{-3mm}
\caption{(a) The correlation between $z$ position of the point with maximum energy deposit and its corresponding drift time. (b) The residual distribution of $z$ position, $z_{project}$ is the $z$ position projected by two regular GEM chambers, $z_{calculate}$ is the $z$ position calculated from the drift velocity and drift time.} \label{thgem_v}
\end{center}
\end{figure}

The electron drift velocity in the gas mixture can be obtained through correlating the hit point's 
 $z$ position with its corresponding drift time in the THGEM chamber. The hit point with maximum energy deposit of a large zenith angle track is selected firstly. The $x$, $y$ positions ($x_{1}$, $y_{1}$) of this point are calculated by the center of gravity method. The $z$ position ($z_{1}$) of this point is then derived through the equation  $ \frac{\sqrt{(x_{0}-x_{2})^{2}+(y_{0}-y_{2})^{2}}}{|z_{0}-z_{2}|} = \frac{\sqrt{(x_{0}-x_{1})^{2}+(y_{0}-y_{1})^{2}}}{|z_{0}-z_{1}|}$,  where $x_{0}$, $x_{2}$, $y_{0}$, $y_{2}$, $z_{0}$, $z_{2}$ have been defined in the previous section. This process is depicted in Fig.~\ref{driftv_principle}. The correlation between $z$ position and its corresponding drift time is shown in Fig.~\ref{thgem_v}(a). The electron drift velocity can then be extracted from a linear fit to the correlation. The electron drift velocity (at drift E $\approx$ 0.26 kV/cm) obtained using this method is 2.26 cm/$\mu$s, consistent with the data commonly used in the literature~\cite{driftv_data}. The spatial resolution of the THGEM chamber in $z$ direction can be derived by plotting the difference of $z$ position projected by two regular GEM chambers and the $z$ position calculated from the drift velocity and drift time. The overall spatial resolution in $z$ direction is $\sim$1.1 mm, as Fig.~\ref{thgem_v}(b) depicts, including the uncertainty from the trigger clock distribution (TCD) ($\sim$0.2 mm) and the uncertainty from the THGEM and the regular GEM chamber spatial resolutions in $x$, $y$ direction ($\sim$0.8 mm, $\frac{\sqrt{\sigma_{x_{0}}^{2}+\sigma_{x_{1}}^{2}}}{tan\theta}$, this formula is derived under the assumption that the THGEM (GEM) chamber has the same spatial resolution in $x$, $y$ direction. $\sigma_{x_{0}}$ = 0.15 mm is the regular GEM chamber spatial resolution in $x$ direction while $\sigma_{x_{1}}$ = 0.22 mm is the THGEM chamber spatial resolution in $x$ direction, and tan$\theta$ = 0.32 is the average value from Fig.~\ref{thgem_spatial}(a)). With these contributions subtracted, the intrinsic spatial resolution in $z$ direction of the THGEM chamber is $\sim$0.7 mm.
 

\subsection{Track reconstruction capability of the THGEM chamber}
\begin{figure}[htbp]
\begin{center}
\includegraphics[keepaspectratio,width=1.0\textwidth]{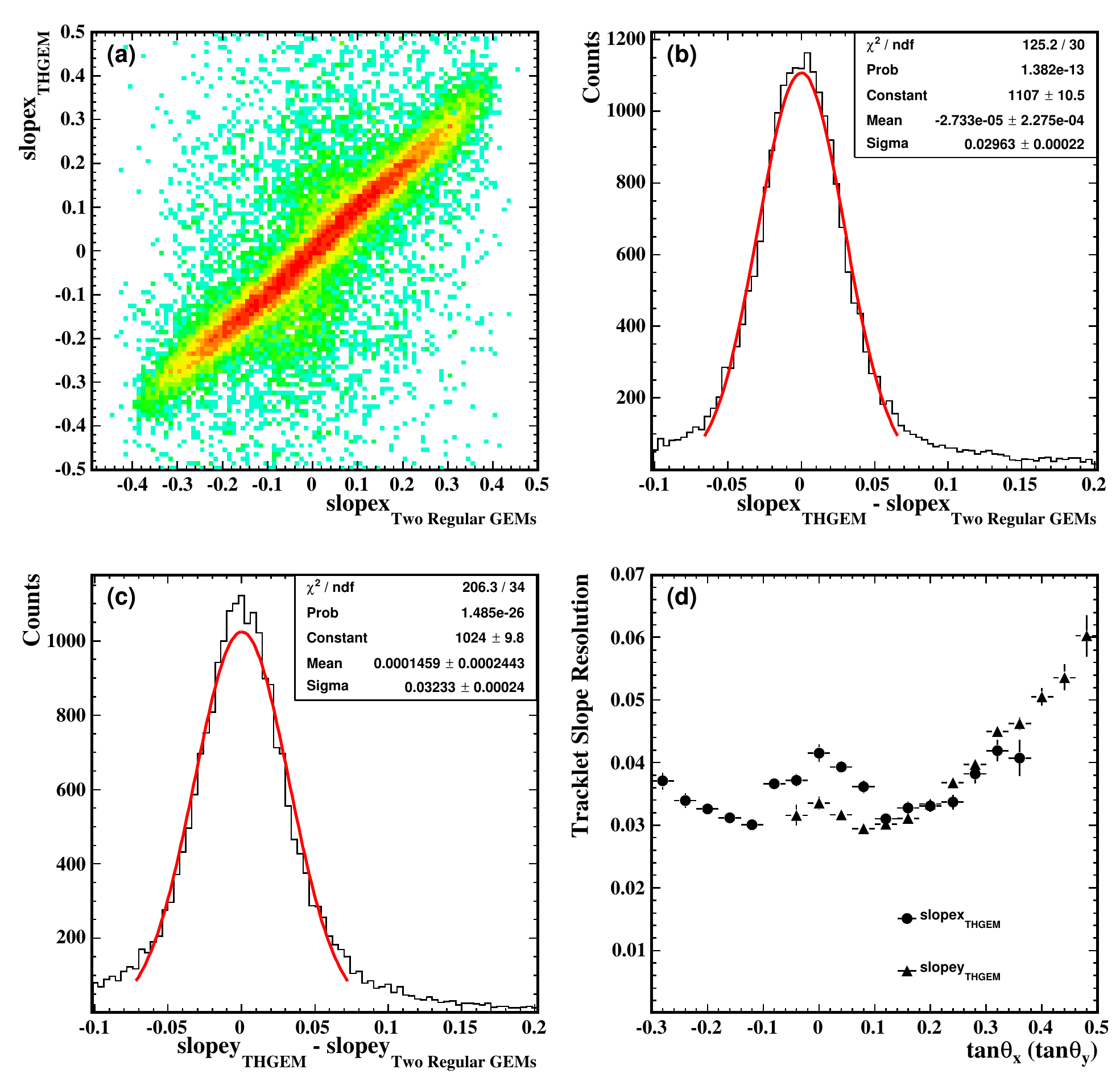}
\vspace*{-3mm}
\caption{(a) The correlation between the tracklet slope measured by the THGEM chamber and that measured by the two regular GEM chambers in $x$ direction. (b) The residual distribution of the tracklet slope in $x$ direction. (c) The residual distribution of  the tracklet slope in $y$ direction. (d) The tracklet slope resolution of the THGEM chamber in $x$ or $y$ direction as a function of incident angle.} \label{track_reconstruction}
\end{center}
\end{figure}

The cosmic ray can be reconstructed just by the THGEM chamber. Firstly, searching for a hit point for each time bucket. If found, the $x$, $y$ positions of this point are calculated using the center of gravity method. Secondly, all these hit points are fitted by a linear function to obtain the cosmic ray tracklet slope (tan$\theta$)  if the number of points is more than two. Thus, the correlation between $slope_{THGEM}$ (measured by the THGEM chamber itself) and $slope_{Two~Regular~GEMs}$ (the slope of the same cosmic ray track measured by the two regular GEM chambers) can be used to study the tracklet slope resolution of the THGEM chamber, which characterizes the track reconstruction capability. The resolution of tracklet slope obtained by the two regular GEM chambers is $\sim$$10^{-3}$ since the distance between the two regular GEMs along $z$ direction is 21.0 cm and the regular GEM's spatial resolution in $x$ or $y$ direction is $\sim$150 $\mu$m. Figure~\ref{track_reconstruction} shows the correlation between tracklet slope measured by the THGEM chamber and that measured by the two regular GEM chambers in $x$ direction, the tracklet slope resolution of the THGEM chamber in $x$ and $y$ direction and the THGEM chamber tracklet slope resolution as a function of cosmic ray incident angle. The tracklet slope resolution in $x$ ($y$) direction is 0.03. Moreover, the results shown in Fig.~\ref{track_reconstruction}(d) indicate that the tracklet slope resolution deteriorates with increasing incident angle as previously observed in similar studies~\cite{spatialResvsAngle}.

\subsection{Gain uniformity and stability of the THGEM chamber}
\begin{figure}[htbp]
\begin{center}
\includegraphics[keepaspectratio,width=1.0\textwidth]{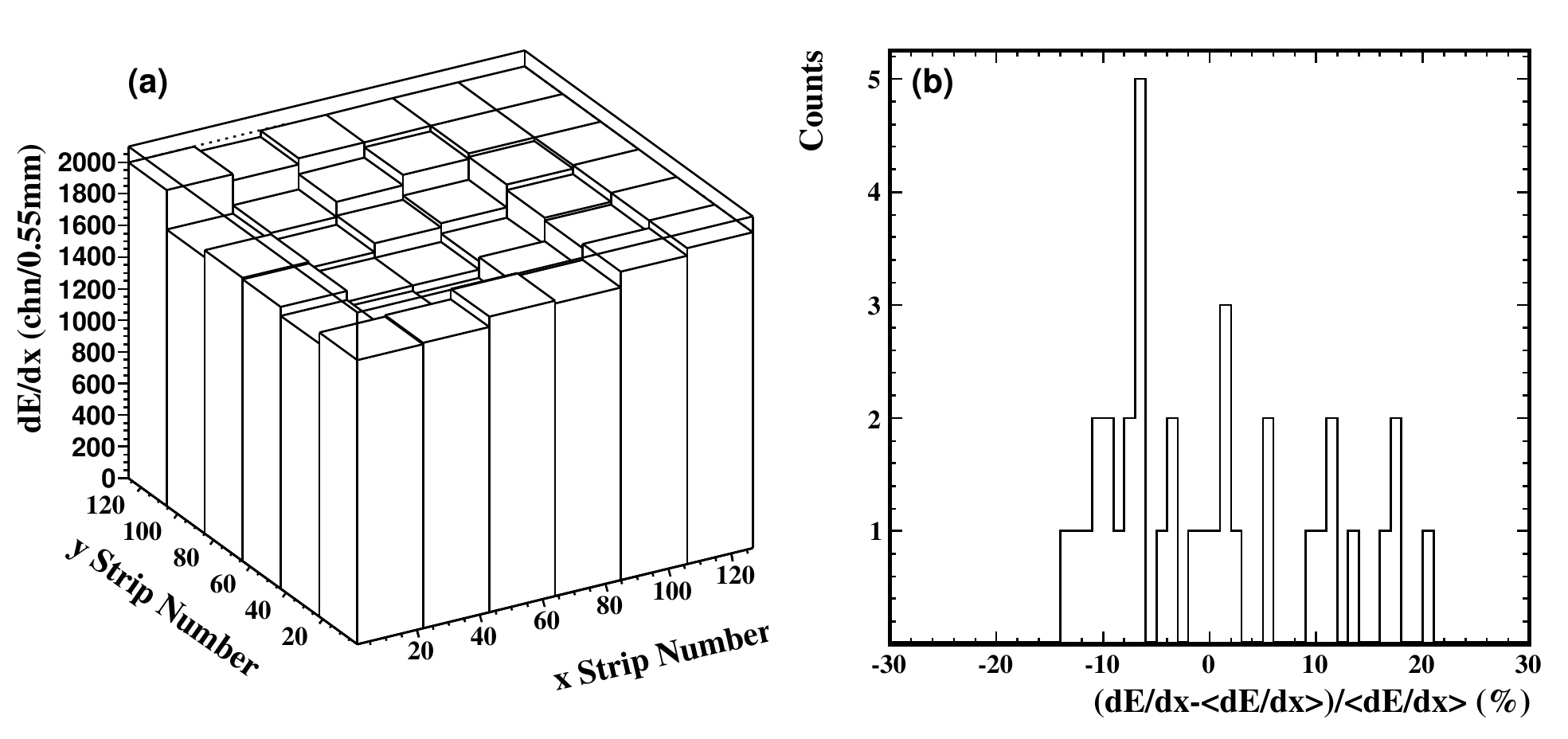}
\vspace*{-3mm}
\caption{(a) \dedx value of each sub-region, \dedx is the average recorded charge for all tracks passing the given region. (b) $dE/dx$ non-uniformity of each sub-region, $<$$dE/dx$$>$ is the average of  \dedx over all regions.} \label{foil_uniformity}
\end{center}
\end{figure}

\begin{figure}[htbp]
\begin{center}
\includegraphics[keepaspectratio,width=0.7\textwidth]{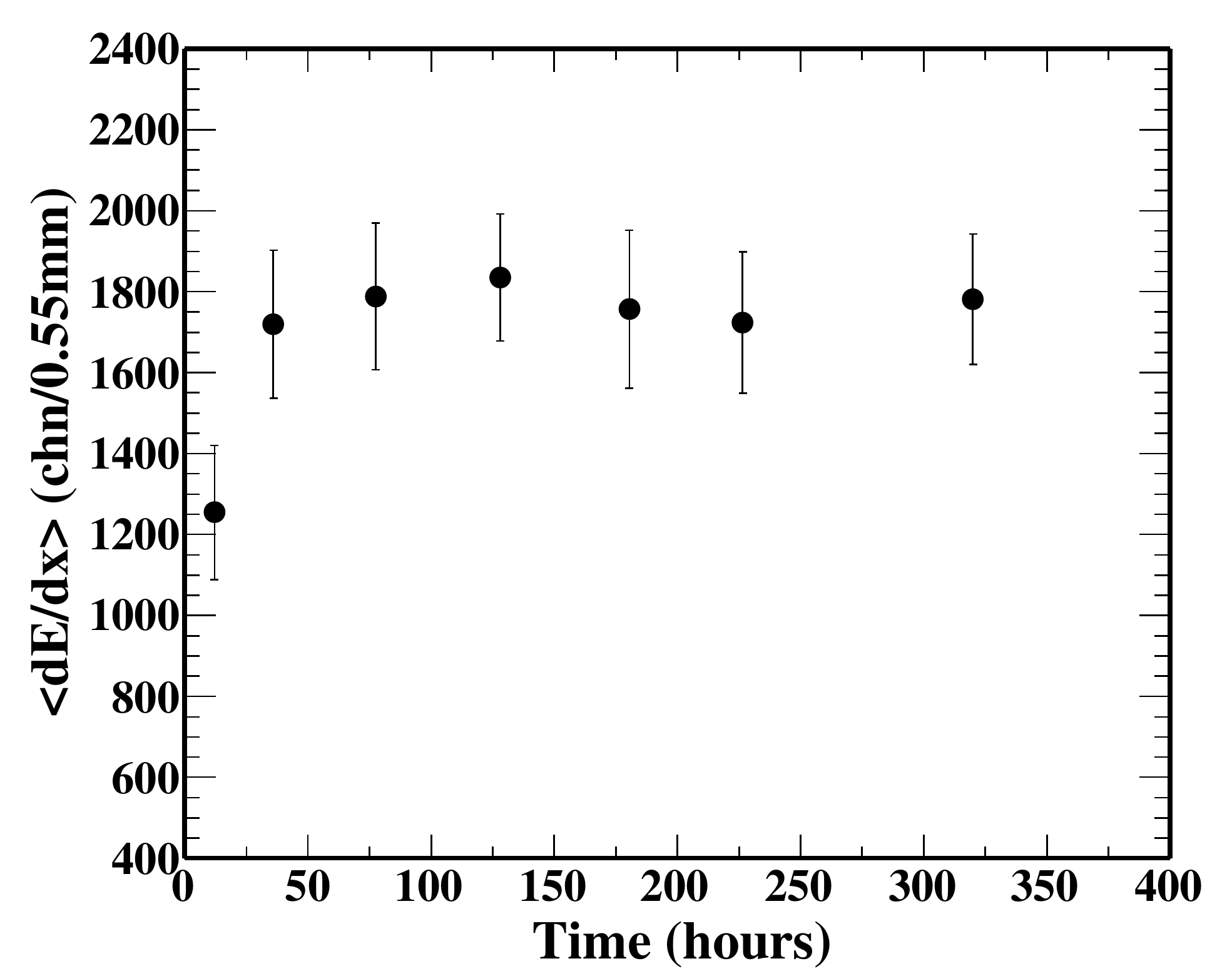}
\vspace*{-3mm}
\caption{The $<$$dE/dx$$>$ as a function of  operating time for the THGEM chamber.} \label{stability}
\end{center}
\end{figure}

The readout board of the THGEM chamber is artificially divided into 6$\times$6 identical sub-regions. The non-uniformity is described using $\frac{dE/dx-<dE/dx>}{<dE/dx>}$, where \dedx is the average recorded charge for all tracks passing the given region while $<$$dE/dx$$>$ is the average of  \dedx over all regions. The measured \dedx of different sub-regions is shown in Fig.~\ref{foil_uniformity}(a). The \dedx non-uniformity for most of the sub-regions (32 out of 36) is less than 15$\%$ as Fig.~\ref{foil_uniformity}(b) depicts. The $<$$dE/dx$$>$ as a function of operating time for the THGEM can be found in Fig.~\ref{stability}. The THGEM shows an increase of gain in the first 24 hours and remains stable afterwards.

\section{Conclusions}\label{conclusion}
A mini-drift THGEM chamber has been tested in laboratory using cosmic rays. The results of the cosmic ray test reveal the excellent performance of the THGEM chamber. The detection efficiency of the THGEM chamber is greater than 94$\%$ and the 1-D spatial resolution is $\sim$220 $\mu$m. Moreover, the THGEM chamber shows good spatial resolution ($\sim$0.7 mm) along electron drift direction and good tracklet slope resolution (0.03) in $x$, $y$ direction, which are important for the small incident angle track reconstruction. In addition to these, the THGEM chamber shows good uniformity and stability during a long run. Such work offers an important reference for the proposed TRD design.

\section{Acknowledgments}
This work was supported in part by (BNL EIC R$\&$D funding) the Nuclear Physics Division within the U.S. DOE Office of Science. This work is supported in part by the National Natural Science Foundation of China under Grant nos. 11375180, 11275196. We express our gratitude to the STAR Collaboration and the RHIC Computing Facility at Brookhaven National Laboratory for their support.



\begin{thebibliography}{00}

\expandafter\ifx\csname
natexlab\endcsname\relax\def\natexlab#1{#1}\fi
\expandafter\ifx\csname bibnamefont\endcsname\relax
  \def\bibnamefont#1{#1}\fi
\expandafter\ifx\csname bibfnamefont\endcsname\relax
  \def\bibfnamefont#1{#1}\fi
\expandafter\ifx\csname citenamefont\endcsname\relax
  \def\citenamefont#1{#1}\fi
\expandafter\ifx\csname url\endcsname\relax
  \def\url#1{\texttt{#1}}\fi
\expandafter\ifx\csname
urlprefix\endcsname\relax\def\urlprefix{URL }\fi
\providecommand{\bibinfo}[2]{#2}
\providecommand{\eprint}[2][]{\url{#2}}

\bibitem{eic} A. Accardi {\it et al.} arXiv:1212.1701 (2012)
\bibitem{erhic} E.C. Aschenauer {\it et al.} arXiv:1409.1633 (2014)
\bibitem{estarloi} The STAR Collaboration, \\ https://drupal.star.bnl.gov/STAR/files/eSTAR-LoI$\_$v30$\_$0.pdf.
\bibitem{startrdproposal} STAR Transition Radiation Detector Proposal: \\https://wiki.bnl.gov/conferences/images/3/3d/TRD$\_$EIC$\_$RDproposal$\_$FY2012$\_$v3.pdf.
\bibitem{tpctofpid} M. Shao {\it et al.}, \Journal{\NIMA}{558}{2006}{419}
\bibitem{alicetrdproposal} ALICE Transition Radiation Detector Proposal: CERN/LHCC 99-13, LHCC/P3-Addendum 2, 7 May 1999.
\bibitem{gem} F. Sauli, \Journal{\NIMA}{386}{1997}{531}.
\bibitem{thgem} R. Chechik {\it et al.}, \Journal{\NIMA}{535}{2004}{303}.
\bibitem{thgem1} A. Breskin {\it et al.}, \Journal{\NIMA}{598}{2009}{107}.
\bibitem{arxethgem} R. Alon {\it et al.}, 2008 JINST 3 P01005.
\bibitem{arxethgem1} J. Miyamoto {\it et al.}, 2010 JINST 5 P05008
\bibitem{arxegem} A. Bondar {\it et al.}, \Journal{\NIMA}{419}{1998}{418}.
\bibitem{arxegem1} A. Bondar {\it et al.}, \Journal{\NIMA}{481}{2002}{200}.
\bibitem{apv} M.J. French {\it et al.}, \Journal{\NIMA}{466}{2001}{359}.
\bibitem{fgtreadout} G.J. Visser {\it et al.}, Real Time Conference (RT), 18th IEEE-NPSS, (2012) p1-6.
\bibitem{gemres} S. Das, RHIC $\&$ AGS Annual User's Meeting, 2013. 
\\https://drupal.star.bnl.gov/STAR/files/gmt$\_$poster$\_$rhicags$\_$21stJune$\_$v2$\_$0.pdf 
\bibitem{driftv_data} T. Zhao {\it et al.},  \Journal{\NIMA}{340}{1994}{485}.
\bibitem{spatialResvsAngle} F. Simon {\it et al.},  \Journal{\NIMA}{598}{2009}{432}.
\end{thebibliography}
\end{document}